\documentclass
[preprint,prl,twocolumn,amsfonts,amssymb,byrevtex,tightenlines,10pt,nobalancelastpage]{revtex4}%
\usepackage{amsfonts}
\usepackage{amsmath}
\usepackage{amssymb}
\usepackage{graphicx}%
\providecommand{\U}[1]{\protect\rule{.1in}{.1in}}

\begin{document}
\preprint{ }
\title{Hot Electron Effects in the 2D Superconductor-Insulator Transition}
\author{Kevin A. Parendo, K. H. Sarwa B. Tan, and A. M. Goldman}
\affiliation{School of Physics and Astronomy, University of Minnesota, Minneapolis, MN 55455}
\pacs{PACS number}

\begin{abstract}
The parallel magnetic field tuned two-dimensional superconductor-insulator
transition has been investigated in ultrathin films of amorphous Bi. The
resistance is found to be independent of temperature on both sides of the
transition below approximately 120 mK. Several observations suggest that this
regime is not intrinsically "metallic" but results from the failure of the
films' electrons to cool. The onset of this temperature-independent regime can
be moved to higher temperatures by either increasing the measuring current or
the level of electromagnetic noise.\ Temperature scaling is successful above
120 mK. Electric field scaling can be mapped onto temperature scaling by
relating the electric fields to elevated electron temperatures. These results
cast doubt on the existence of an intrinsic metallic regime and on the
independent determination of the correlation length and dynamical critical
exponents obtained by combining the results of electric field and temperature scaling.

\end{abstract}
\volumeyear{year}
\volumenumber{number}
\issuenumber{number}
\eid{identifier}
\date[Date text]{date}
\received[Received text]{date}

\revised[Revised text]{date}

\accepted[Accepted text]{date}

\published[Published text]{date}

\startpage{1}
\endpage{4}
\maketitle

Quantum phase transitions (QPTs) \cite{SondhiQPT} have received a lot of
attention as quantum criticality is a ubiquitous feature of many strongly
correlated electron systems. In a QPT, a system flows towards one of two
different ground states as $T\rightarrow0$, depending on the value of a
"tuning" parameter in its Hamiltonian. The nature of the transition\ is
determined by the correlation length and dynamical critical exponents, $\nu$
and $z$ respectively. Non-linear conductivity \cite{SondhiQPT, Phillips,
SondhiNonLinear, Green} near the quantum critical point is important because
the results of the finite size scaling under changes of electric field can be
combined with the scaling under changes of temperature to determine the values
of $\nu$ and $z$. In this process, an important issue is the extent to which
intrinsic non-linear response is more important than Joule heating. While
there is a material dependent criterion \cite{SondhiQPT}, it is based on
dimensional analysis and may thus lack important multiplicative factors.

A second issue is the existence of metallic phases at low temperatures near
the quantum critical point. The Bose-Hubbard model of the
superconductor-insulator (SI) transition in two dimensions predicts a
metallic\ ground state having finite resistance at zero temperature only at
the critical value of the tuning parameter \cite{Fisher}; slight deviations
from criticality result in either superconducting or insulating ground states.
However, as measurements have been extended to lower temperatures,
temperature-independent resistances have been found on both sides of the
transition over an extended range of tuning parameters \cite{Yazdani, Mason,
ParendoPRB, Yoon}. There have been a couple of claims that these are evidence
of an intrinsic metallic regime between the superconducting and insulating
states \cite{Mason, Yoon}, which have led to considerable discussion of
reformulations of the SI\ transition scenario \cite{Das&Doniach,
Phillips&Dalidivich, Kapitulnik, Galitski}. The nature of this resistance
saturation is far from certain because of difficulties with electrical
measurements on ultrathin films at mK temperatures. The electronic heat
capacities of such films are very small, so that even modest levels of
dissipation may prevent cooling of the films' electrons. Dissipation can be
due to either the measuring current or from currents that result from the
electromagnetic noise environment. At low temperatures, the dominant cooling
mechanism for electrons is through the electron-phonon interaction, and this
coupling is weak at mK temperatures \cite{Clarke}. A refrigerator could thus
cool a film's phonons more effectively than its electrons, resulting in a
temperature-independent resistance since the film would be measured at its
minimum achievable electron temperature over an extended range of lower phonon
or lattice temperatures.

We discuss here the role of electron heating in the 2D SI\ transition in
ultrathin films. An electric field scaling analysis that successfully
collapses data is shown to be a direct consequence of heating and not due to
quantum critical non-linear electrical response. Since electric field scaling
can be mapped onto temperature scaling by relating electric fields to elevated
temperatures, the separate determination of $\nu$ and $z$ is an open issue.
Also, the temperature-independent resistance below 120 mK that has previously
been asserted for other superconducting films to be evidence of a novel
metallic state is likely a consequence of heating due to the electromagnetic
environment. This regime of temperature-independent resistance prevents the
extension of temperature scaling below 120 mK. The relevance of these
observations to other 2D SI\ transitions is discussed.

A 10.4 \AA \ thick \textit{a}-Bi film was deposited onto 10 \AA \ of amorphous
antimony (\textit{a}-Sb) that was pre-deposited onto a SrTiO$_{3}$ substrate
that was held at 6 K during depositions. Such films are believed to be
disordered on an atomic, rather than mesoscopic, length scale; recent
structural studies support this hypothesis \cite{Footnote}. The film was then
transferred to a dilution refrigerator without removing the sample from vacuum
or warming it above 15 K \cite{Hernandez}. The system was heavily shielded and
electrically filtered to minimize the electromagnetic noise in the film's
environment \cite{ParendoPRB2} with AC filters at 300 K to remove noise at 60
Hz and at radio frequencies and at mK temperatures to remove noise at GHz
frequencies. The sheet resistance, $R$, and the differential sheet resistance,
$R_{D}=dV/dI$, were measured by applying a DC current, $I$, across the film
and measuring voltage, $V$, across an area of $(0.5mm)^{2}$ of film.
Resistance was determined from $(V(I)-V(-I))/(2I)$, with $I=1$ $nA$. $R_{D}$
was determined from $V((I+\Delta I)-V(I-\Delta I))/2\Delta I$ with $\Delta
I=1$ nA, with $I=$ 0, 5, 10, 25, 35, 50, 65, 80, 100, and 500 nA.

Magnetic fields, $B$, applied parallel to the film plane and perpendicular to
the direction of the measuring current, were used to tune the transition. In
Fig. 1(a), $R(T)$ is shown for several values of $B$. At $B=0$ T, $T_{c}$ was
about 80 mK. At 12 T, $R(T)$\ was best described by 2D Mott variable range
hopping from 120 mK to 10 K. The $R(T,B)$ data were then analyzed using
temperature scaling. Note that data at temperatures below 120 mK were excluded
from the analysis as below 120 mK $R(T)$ deviated from its high temperature
behavior and became independent of temperature. For isotherms between 120 and
250 mK, for $B$ between 3 and 12 T, values of $R(B)$ crossed at a critical
resistance $R_{c}=$ 15,200 $%
\Omega
$ and a critical field $B_{c}=6.87$ T. In Fig. 1(b), $R/R_{c}$ is plotted
against the temperature scaling parameter $|B-B_{c}|T^{-1/\nu z}$ with the
value of $\nu z=0.68$ $\pm$ $0.05$ that produced the best collapse of data.
This value of $\nu z$ agrees with those found when perpendicular fields
\cite{Markovic}, electrostatic charge transfer \cite{ParendoPRB2}, and
parallel fields \cite{ParendoPRB2} were used as tuning parameters for
\textit{a}-Bi films.%

\begin{figure}
[ptb]
\begin{center}
\includegraphics[
height=2.8443in,
width=2.919in
]%
{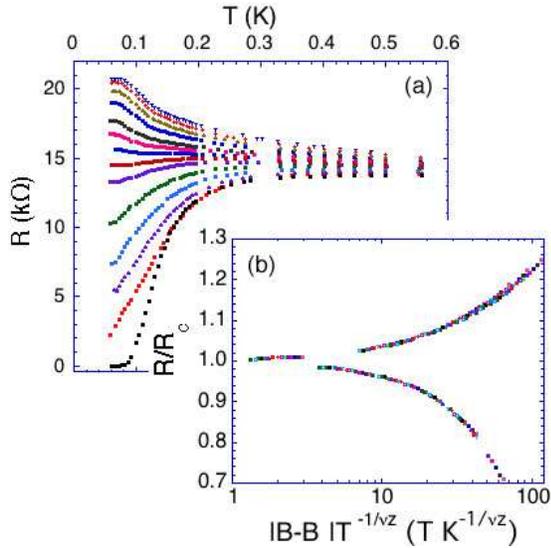}%
\caption{(a) Resistance vs. temperature at $B=$ 0 (bottom), 2, 3, 4, 5, 6,
6.5, 7, 7.5, 8, 9, 10, 11, and 12 T (top). (b) Temperature scaling for this
transition.}%
\end{center}
\end{figure}

Data were collected for electric field scaling by recording $R_{D}$ at 10
values of bias current at the same $B$ fields and temperatures as for the
temperature scaling. In Fig. 2(a), we show $R_{D}$ vs. $I$ recorded at 65 mK.
For this data, curves of $R_{D}$ vs. $B$ at different values of $I$ crossed at
the same values of $R_{D,c}$ and $B_{c}$ as in temperature scaling, within
uncertainty. We show the electric field scaling analysis in Fig. 2(b), where
$R_{D}/R_{D,c}$ is plotted against the electric field scaling function
$|B-B_{c}|V^{-1/\nu(z+1)}$ for the value of $\nu(z+1)=2.0$ $\pm$ $0.1$ that
produced the best collapse of data. The two scaling analyses suggest that $\nu
z\sim0.68$ and $\nu(z+1)\sim2.0$, which then yield $\nu\sim1.3$ and $z\sim
0.5$. However, this value of $z$ is unphysical as $z$ is believed to be either
1 or 2 for charged Bose systems, depending on their interactions \cite{Herbut}.

We now show that the currents in the electric field scaling analysis heat the
electrons. Each value of current corresponds to an elevated electron
temperature. When this is taken into account, the collapse in electric field
scaling can be shown to be due to temperature scaling.%

\begin{figure}
[ptb]
\begin{center}
\includegraphics[
height=2.8701in,
width=2.919in
]%
{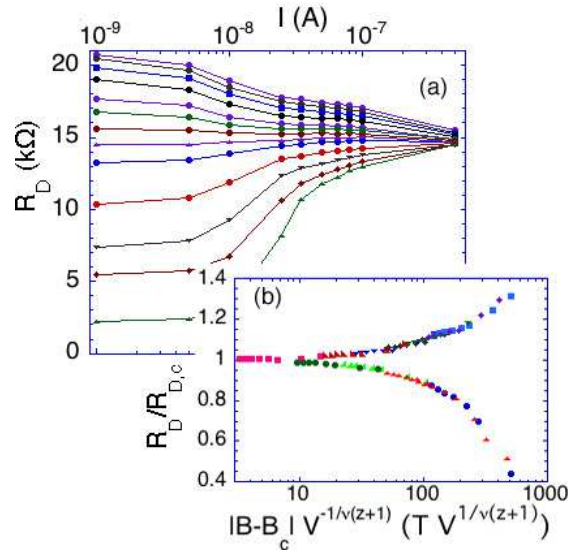}%
\caption{(a) Differential resistance vs. current at 65 mK for $B$ = 2
(bottom), 3, 4, 5, 6, 6.5, 7, 7.5, 8, 9, 10, 11, and 12 T (top). (b) Electric
field scaling for this transition.}%
\end{center}
\end{figure}

Some evidence for heating by the measuring current is seen in the temperature
dependence of $R_{D}$. In Figs. 3(a)-3(d), we show curves of $R_{D}(T)$ for
magnetic fields between 0 and 12 T at four illustrative bias currents, 0, 25,
65, and 100 nA. Above 300 mK, $R_{D}(T)$ curves were independent of $I$ for
given value of B. However, below 300 mK, $R_{D}$ was independent of
temperature at high $I$. Indeed, with $I=100$ nA, $R_{D}$ was independent of
temperature up to almost 200 mK, which is highly suggestive that this current
heated the electrons as high as 200 mK for all refrigerator temperatures below
200 mK.%

\begin{figure}
[ptb]
\begin{center}
\includegraphics[
height=3.4761in,
width=3.4006in
]%
{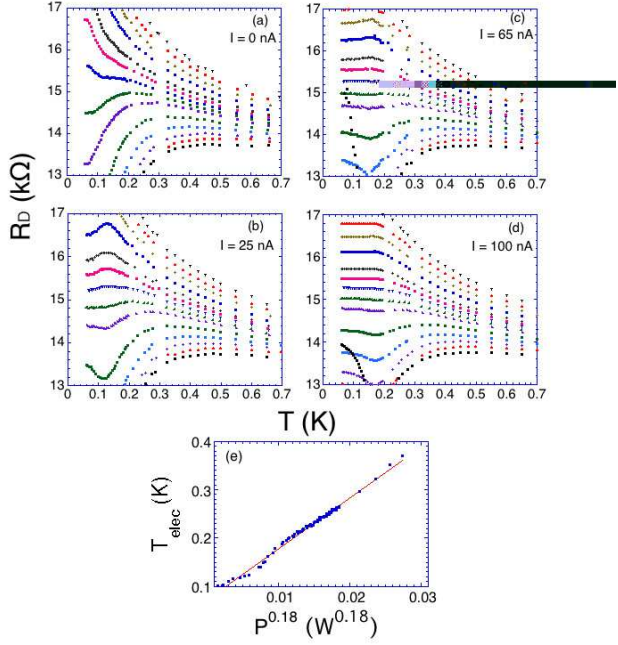}%
\caption{Differential resistance vs. temperature at 0 (bottom), 2, 3, 4, 5, 6,
6.5, 7,7.5, 8, 9, 10, 11, and 12 T (top), for four measurement currents. (e)
$T_{elec}$ vs. measurement power.}%
\end{center}
\end{figure}

We now explore the consequences of the hypothesis that electron heating is the
cause of deviations of $R_{D}$ from $R$ at given B and T. We use $R(T)$ as a
thermometer, so that each value of resistance corresponds to an electron
temperature. Keeping the refrigerator at 65 mK, each current produces a value
of $R_{D}$ that corresponds to a higher effective electron temperature,
$T_{elec}$. Thus, we assume that\ the only effect of increased current is to
heat the film's electrons. The effective electron temperature is found to
increase with power as $T_{elec}\sim P^{0.18}$ for $I\geq5$ nA which is shown
in Fig. 3(e) for a 7.89 $\mathring{A}$\ thick film of a similar sample, where
more data were available from more extensive measurements of $I-V$
characteristics. This same exponent was found, within $\pm0.05$, for the
$10.4$ $\mathring{A}$\ thick film, where less data were available. This power
dependence is very close to that proposed by Wellstood \textit{et al}.
\cite{Clarke} ($T_{min}\sim P^{1/5}$) to describe the relationship between a
metal film's minimum electron temperature and measurement power. They verified
this relationship experimentally by varying the bias power in AuCu films and
determining the electron temperature using noise thermometry.

We can now map electric field scaling onto temperature scaling. On the
superconducting side of the transition, $I-V$ characteristics are
Josephson-like, while on the insulating side, they are single-particle like.
Near $B_{c}$, they are only slightly non-linear. Over the current range used
for electric field scaling ($I\geq5nA$), $V\sim I^{1.05}$ at $3$ $T$, while
$V\sim I^{0.95}$ at $12$ $T$. The average effect is roughly $I\sim V$. Since
$P=IV$, then $T_{elec}\sim V^{0.36}$, or $V\sim T_{elec}^{2.78}$. If this
relationship for $V$, which is proportional to the electric field, is inserted
into the electric field scaling relation, we find $R_{D}/R_{D,c}%
=F(|B-B_{c}|T_{elec}^{-1/0.72})$. This value of $\nu z=0.72$ agrees within
error with that found by temperature scaling. This suggests that the electric
field scaling works because it is effectively temperature scaling, and
increased measuring current increases temperature. There is a slight
uncertainty in this result since $I$ vs. $V$\ is slightly different for each
value of B. However, when data used for electric field scaling are analyzed
using temperature scaling by determining the electron temperature using $R(T)$
as a thermometer, the resulting analysis collapses on top of the temperature
scaling analysis from Fig. 2. The exponent product and scaling function are
\textit{identical} for electrons warmed by these two methods. This is shown in
Fig. 4.%

\begin{figure}
[ptb]
\begin{center}
\includegraphics[
height=2.3495in,
width=3.1598in
]%
{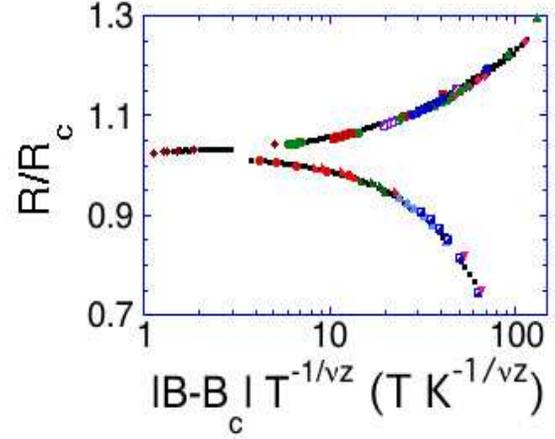}%
\caption{Temperature scaling for data obtained by raising refrigerator
temperature (small black circles) and for data obtained by heating the
electrons with elevated measurement currents (larger color symbols).}%
\end{center}
\end{figure}

Thus, for \textit{a}-Bi films, electric field scaling cannot provide
information to separately determine $\nu$ and $z$. Is this merely a property
of the \textit{a}-Bi/\textit{a}-Sb film, or is it a more general result? In
the limit of zero temperature, non-linear transport effects are expected to
compete with Joule heating, with material specific properties being important.
The electron temperature varies with power as $T_{elec}\sim P^{1/\theta}$ with
$\theta=p+2$, where p is the temperature coefficient of the inelastic
electron-phonon scattering rate, $\tau_{in}^{-1}\sim T^{p}$ \cite{Bozler}.
There is a criterion \cite{SondhiQPT} that for $\frac{2}{\theta}<\frac{z}%
{z+1}$ Joule heating, rather than intrinsic non-linear effects will dominate.
This criteria arises from a dimensional analysis argument that may ignore
important multiplicative factors. We do not know z, but assuming it to be 1 or
2, the value of $\theta$ in \textit{a}-Bi meets this criterion as stated.
However, even in the earlier work of Yazdani and Kapitulnik on MoGe thin films
\cite{Yazdani}, in which scaling analyses revealed apparent values of $\nu
z\sim1.35$ and $\nu(z+1)\sim2.65$, which yielded $\nu\sim1.3$ and $z\sim1$,
heating may actually be responsible. It is known that $p=2$ in MoGe
\cite{Graybeal}, so that $\theta=4$. This puts MoGe in the "marginally
dangerous" category in which both heating and intrinsic effects may be
important \cite{SondhiQPT}. Indeed, this value of $\theta$ implies that $V\sim
T^{2}$, implying that in the electric field scaling of Yazdani and Kapitulnik,
$V^{-1/2.65}\sim T_{elec}^{-2/2.65}=T_{elec}^{-1/1.375}.$ Thus, electric field
scaling maps well onto temperature scaling if one assumes heating is
responsible for the electric field dependence.

We note similarities of our results with those of Golubkov \textit{et al.
}\cite{Golubkov} in In$_{2}$O$_{3}.$ They suggested that electron heating was
dominant on the insulating side of the perpendicular field tuned transition
with $\theta=5$. However, they found that data collapse in an electric field
scaling analysis was "unsuccessful." This may imply that vortex motion is a
more dominant effect than heating on the superconducting side of their transition.

Indeed, most dirty samples have values of p between 2 and 4 \cite{Ovadyahu}
raising the possibility that heating will dominate intrinsic non-linear
effects in most materials.

Given that elevated current can cause Joule heating in these films, what
happens in measurements made with minimal bias currents? A second cause of
heating is the current induced in a film by its electromagnetic environment.
In a similar series of films, we measured $R(T)$ before and after removal of
an AC filter that removed noise at 60 Hz. $R(T)$ was unchanged above 150 mK.
The temperature below which $R$ became independent of temperature moved to
higher $T$ after the removal of the filter. Since the DC\ resistance cannot
change in any way by the removal of an AC\ filter, this must be caused by
heating of electrons in the film due to increased noise current. It is
reasonable to assume that even with very strong filtering, the
residual\ electromagnetic noise caused $R(T)$ to become independent of
temperature as $T\rightarrow0$ as the electron-phonon coupling weakens.

One might ask how general is the above conclusion. The extent of the shielding
and filtering in the measurements on nominally homogeneous MoGe and Ta films
used as the basis for claims of a low T metallic regime is not revealed in
Refs. 7 and 9. Measurements of mesoscopically clustered films show minima in
$R(T)$ in $B=0$ at the bulk transition temperature, while the resistance is
independent of temperature at temperatures below these minima \cite{Jaeger}.
It is possible that these films exhibit intrinsic metallic phases at low
temperatures, as the temperature independence extends to temperatures on the
order of several K, in a regime where electron-phonon coupling and other heat
transfer mechanisms should be strong. On the other hand, it is possible that
these films act as better antennae for\ radiation or that the clusters change
the mechanism of heat transfer in an unknown manner. A final caveat is that
although resistance saturation for parallel-field, perpendicular field,
electrostatic, and thickness tuned transitions are qualitatively similar, the
perpendicular field case involves vortices and the physics may be different.

In conclusion, we have found that measuring currents in ultrathin
\textit{a}-Bi/\textit{a}-Sb films can heat electrons out of equilibrium with
their environment. This prevents electric field scaling from yielding
information to separately determine $\nu$ and $z$. Residual electromagnetic
noise and non-zero measuring currents may lead to the so-called metallic
regime observed in a number of experiments at temperatures below about 150 mK.

The authors would like to thank Professors H. Suhl and B. Shklovskii for
useful discussions. This work was supported by the National Science Foundation
under grant NSF/DMR-0455121.

\end{document}